\def\year{2022}\relax
\documentclass[letterpaper]{article} 
\usepackage{aaai21}  
\usepackage{times}  
\usepackage{helvet} 
\usepackage{courier}  
\usepackage[hyphens]{url}  
\usepackage{graphicx} 
\usepackage{natbib}  
\usepackage{caption} 
\usepackage{booktabs} 
\usepackage{amsfonts}
\usepackage{amsmath}
\usepackage{graphicx}
\usepackage{paralist}
\usepackage{balance}
\usepackage[T1]{fontenc}
\usepackage{subcaption}
\usepackage{algorithm2e}
\usepackage{amsthm}
\usepackage{hyperref}
\usepackage{xcolor}
\usepackage{pifont}
\usepackage{multirow}

\usepackage{mathtools}
\usepackage{listings} 
\usepackage{bbm}
\usepackage{adjustbox}
\usepackage{array}
\usepackage{enumitem}

\newcommand{\spara}[1]{\smallskip\noindent{\bf #1}}

\newcommand{\rurl}[1]{\href{http://#1}{#1}}

\title{SciLander: Mapping the Scientific News Landscape}

\author{}
\author {
    Maur\'{i}cio Gruppi,\textsuperscript{\rm 1}
    Panayiotis Smeros, \textsuperscript{\rm 2}
    Sibel Adal\i \textsuperscript{\rm 1} 
    Carlos Castillo \textsuperscript{\rm 3} 
    Karl Aberer \textsuperscript{\rm 2}
    \\
}
\affiliations {
    \textsuperscript{\rm 1} Rensselaer Polytechnic Institute, USA \\
    \textsuperscript{\rm 2} École Polytechnique Fédérale de Lausanne (EPFL), Switzerland \\
    \textsuperscript{\rm 3} Universitat Pompeu Fabra, Spain \\
    gouvem@rpi.edu, panayiotis.smeros@epfl.ch, adalis@cs.rpi.edu, chato@acm.org, karl.aberer@epfl.ch
}

\begin{document}
\maketitle
\begin{abstract}
The COVID-19 pandemic has fueled the spread of misinformation on social media and the Web as a whole.
The phenomenon dubbed `infodemic' has taken the challenges of information veracity and trust to new heights by massively introducing seemingly scientific and technical elements into misleading content.
Despite the existing body of work on modeling and predicting misinformation, the coverage of very complex scientific topics with inherent uncertainty and an evolving set of findings, such as COVID-19, provides many new challenges that are not easily solved by existing tools. 
To address these issues, we introduce SciLander, a method for learning representations of news sources reporting on science-based topics.
SciLander extracts four heterogeneous indicators for the news sources; two generic indicators that capture (1) the copying of news stories between sources, and (2) the use of the same terms to mean different things (i.e., the semantic shift of terms), and two scientific indicators that capture (1) the usage of jargon and (2) the stance towards specific citations.
We use these indicators as signals of source agreement, sampling pairs of positive (similar) and negative (dissimilar) samples, and combine them in a unified framework to train unsupervised news source embeddings with a triplet margin loss objective.
We evaluate our method on a novel COVID-19 dataset containing nearly 1M news articles from 500 sources spanning a period of 18 months since the beginning of the pandemic in 2020.
Our results show that the features learned by our model outperform state-of-the-art baseline methods on the task of news veracity classification.
Furthermore, a clustering analysis suggests that the learned representations encode information about the reliability, political leaning, and partisanship bias of these sources.
\end{abstract}





\section{Introduction}\label{sec:introduction}

\begin{figure}
    \centering
    \includegraphics[width=.8\linewidth]{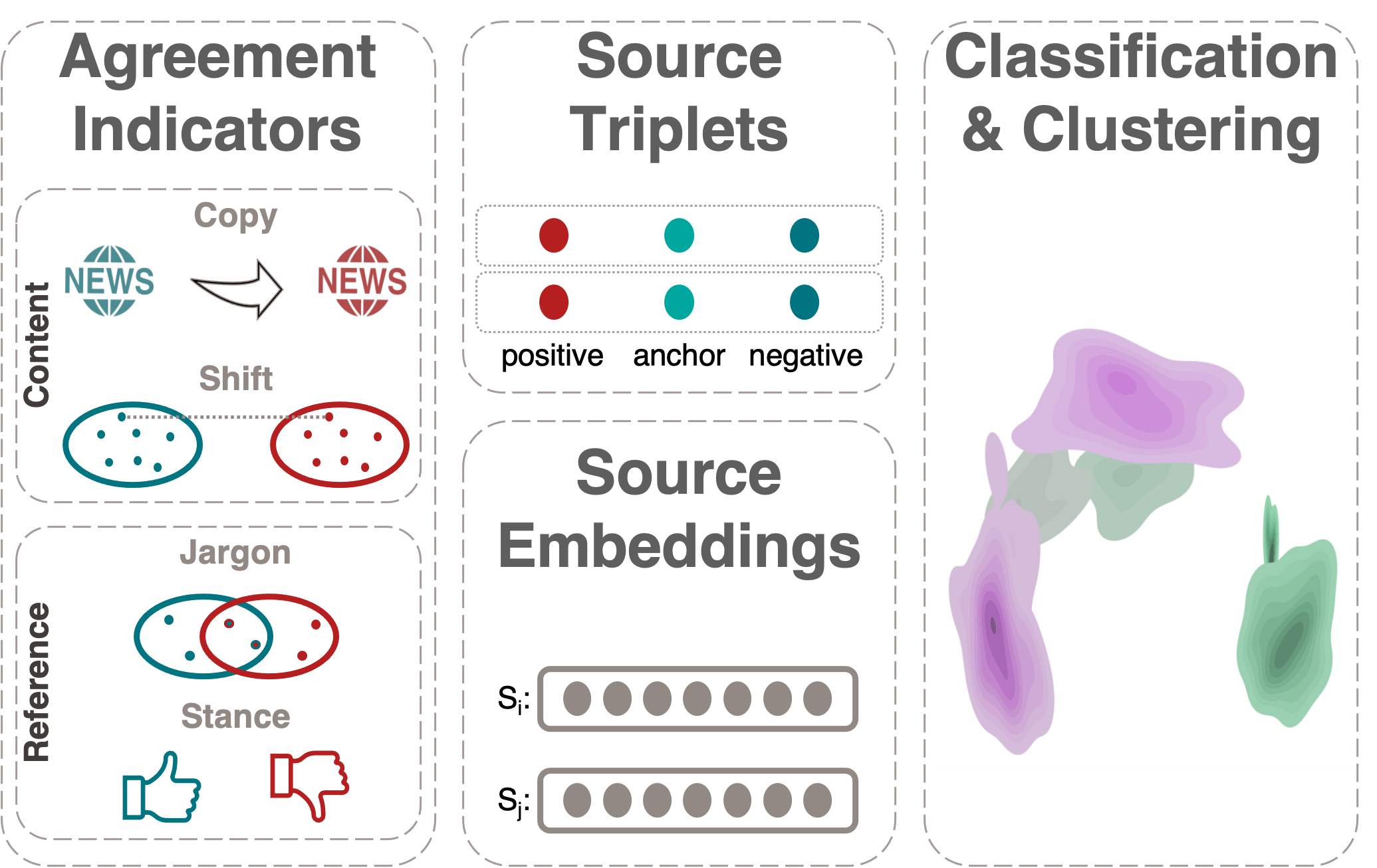}
    \caption{Overview of SciLander, including agreement indicator extraction (\S\ref{sec:content-indicators} \& \S\ref{sec:reference-indicators}), triplet sampling and unsupervised source embeddings training (\S\ref{sec:methods-ensemble}), and evaluation on the downstream tasks of classification and clustering (\S\ref{sec:experiments}).} 
    \label{fig:SciLander_Overview}
\end{figure}

The COVID-19 pandemic has resulted in a significant increase in information
production and consumption at the same time. 
With this came a large increase in unreliable information, dubbed `infodemic' \cite{buchanan2020managing}. 
This increase was also coupled with the growing scrutiny of media sources
and purposeful amplification of any errors they made. 
As the readers sought correct, timely, and trustworthy information, many news and media sources worked hard to discredit others and create confusion \cite{Nature:Bavel2020}.

Governments and public health agencies have the responsibility to respond to the crisis and protect the public from misinformation by utilizing the power of social and news media \cite{castillo2016big}.
Yet, the same social and news media work as a catalyst for the infodemic, allowing disinformation to be dispersed on a large scale, regardless of the significant effort to hinder its spread \cite{doi:10.1177/1065912920938143}.

Despite the existing body of work on modeling and predicting misinformation, coverage of a  complex scientific topic with inherent uncertainty and evolving set of findings, such as COVID-19, provides many new challenges that are not easily solved by existing tools \cite{zarocostas2020fight}.
On the article level, the evaluation of news stories may be challenging as they may contain information that cannot be easily verified. 
Furthermore, many sources may not have the necessary staffing for the proper communication of science-related topics, they may be known to have published incorrect information, this information may also have changed over time, or the source may have later corrected it.

Often, language-based methods fail in such a task because different sources may use the same terms to mean different things. 
Furthermore, many sources may use scientific references to back up their claims; however, the validity of these references is not easily verifiable.
Being able to map out the consequential and systematic patterns of behavior of such sources in terms of both \emph{content} and \emph{references} would be particularly useful in such scenarios \cite{10.1111/j.1083-6101.2011.01565.x}. 
It would allow sources to be compared to other known sources in terms of their coverage, and develop explanations to the aspects in which they are similar to or different from each other. 

To address these challenges, we introduce a novel method called SciLander. 
SciLander builds on a set of novel features, based on the deep processing of news articles published by a set of sources, 
producing a vector representation of these news sources. 
To build this, we incorporate measures of similarity and difference between the sources based on their citation behavior, the republishing of articles from each other, and their general language usage.
In particular, we use the coverage of COVID-19 to show that this embedding has many desirable features that can help multiple downstream tasks.

\spara{Our Contribution.}
%
%
The technical contributions we introduce are the following:
\begin{itemize}[leftmargin=*]
\item We propose four news agreement indicators for sources:
    \begin{inparaenum}[i)]
    \item the shared content or republished articles,
    \item the semantic shift of terms in the common vocabulary,
    \item the usage of scientific jargon, and
    \item the citation stance of the news sources (\S\ref{sec:content-indicators} \& \S\ref{sec:reference-indicators});
    \end{inparaenum}
\item We combine these indicators in a unified framework for training unsupervised news source embeddings (\S\ref{sec:methods-ensemble});
\item We evaluate our method using a dataset of news publications related to COVID-19. Sources in this dataset are labeled with respect to reliability and political leaning;
\item We compare our method to strong baselines on the problem of veracity classification of news sources and show a significant gain in performance when combining the indicators proposed in this paper;
\item We test the applicability of our method in an online learning experiment, showing that it can be used to learn features from sources even if little data is available or if new coming sources are presented in the landscape;
\item We show that the learned features encode information about the sources' reliability level, partisanship bias, and political leaning through a clustering analysis experiment.
\end{itemize}

\section{Related Work}\label{sec:related_work}
%
We distinguish three levels of granularity for misinformation in news and social media: claims, articles, and sources.
This is a broad research area where results are scattered through multiple disciplines and venues; below we present studies relevant to each of the three aforementioned levels.

\spara{Claims \& Articles Veracity.}
Many of the computational methods for veracity assessment of news articles employ machine learning techniques
in supervised binary or multi-label classification settings 
\cite{DBLP:conf/naacl/BalyKSGN19, DBLP:journals/expert/ReisCMVBC19, DBLP:conf/www/ZhangRMASGAVLRB18, DBLP:conf/aaai/YangSWG0019, DBLP:journals/cogsr/VishwakarmaVY19}.
In this setting, a news article is given as the input to a model and it must predict whether the article contains false information.
Other studies aim at detecting the veracity of information at a more granular level by working with claims and rumors 
\cite{DBLP:journals/ipm/ZubiagaKLPLBCA18, DBLP:conf/acl/ShaarBMN20, DBLP:conf/www/HansenHASL19, DBLP:conf/www/JiangBI020, DBLP:conf/cikm/Smeros0A21}.
This approach consists in detecting fragments of text, e.g., sentences or paragraphs, worthy of fact-checking. Thus, a single document or news article may contain several claims, some of which may be inaccurate or deceiving.



\spara{Source Veracity.}
Source-based approaches are holistic approaches that evaluate the quality of a news source as a whole, without focusing on individual claims or articles extracted from it.
%
%
\citet{DBLP:conf/emnlp/BalyKAGN18, DBLP:conf/naacl/BalyKSGN19, DBLP:conf/acl/LiG19} highlight the importance of features beyond text to evaluate the veracity of news sources, such as the presence in social media and the existence of a Wikipedia page about a source.
Furthermore, \citet{DBLP:conf/wsdm/ShuWL19} explore the interactions between users, authors, and sources, while 
\citet{DBLP:journals/corr/abs-2101-10973} observe content sharing trends among news publishers.
Finally, \citet{DBLP:conf/www/BourgeoisRA18, DBLP:conf/www/RappazBA19} study the selection bias in the topic coverage of news sources by exploring the co-references of these sources to the same news events, while
\citet{DBLP:conf/icwsm/RibeiroHBCKBG18} infer the biases of news sources by utilizing their advertiser insights into the demographics of their social media audience.

Both claim- and article-level veracity assessments require data labeling at a very large scale (e.g., individual claims or articles labeled as \emph{reliable} or \emph{unreliable}) and heavily rely on text-specific features these short pieces of text provide.
Our approach is, to the best of our knowledge, the first approach that aggregates information about the writing style and citation behavior of news sources to learn unsupervised source representations, that is aware of the science-related content published by them.
\section{Corpus}\label{sec:corpus}

Our study targets the reliability of sources 
when reporting news related to science.
Thus, we use a corpus of news articles targeted on the emerging scientific topic of COVID-19, and a corpus of scientific references, also targeted on COVID-19. We summarize the basic statistics of both corpora in Table~\ref{tab:corpus-stats}. 

\spara{NELA-GT-2020.}
The collection of news articles contains a total of $1.78$ million articles published by $519$ sources \cite{DBLP:journals/corr/abs-2102-04567}.
Each article in the dataset contains a title, full text, name of the publishing source, and publication timestamp.
We use a subset containing only articles related to COVID-19, resulting in $991{,}116$ news articles from $493$ sources, published over 18 months, between January 1st 2020 and July 1st 2021.
We obtain this subset by applying keyword-based filtering using the COVID-19 terminology from Shugars et al. \cite{shugars2021pandemics}, selecting articles that contain at least one COVID-related keyword in the title or body text.

\spara{Media Bias/Fact Check Labels.}
We retrieve labels for sources in the corpus from the news assessment agency Media Bias/Fact Check\footnote{\url{https://mediabiasfactcheck.com}}.
We obtain the \emph{political leaning} of news sources, represented by direction (left or right) and magnitude (mild, moderate, extreme).
These are encoded as integer numbers in $[-3,3]$, negative values indicate left-bias, positive values indicate right bias, and $0$ represent center sources.
Furthermore, we obtain a \emph{conspiracy-theory} label, a binary indicator denoting whether a source publishes conspiracy theories and/or pseudoscience content.
These are often highly unreliable sources and may or may not exhibit political leaning.
Finally, we obtain \emph{factual reporting}, an integer score from 0 to 5 assigned to each source, where 0 indicates the least credible score and 5 is the most credible score. A source that constantly publishes misleading content, fails to fact-check its publications, and does not disclose an editorial board tends to be associated with a lower factual reporting score.

Based on the \emph{factual reporting} score, we divide news sources into two reliability classes, namely the \emph{Reliable News Sources} and the \emph{Uneliable News Sources}. The rules defining each class are described as follows:
\begin{itemize}
    \item \emph{Reliable News Sources}: sources whose \emph{factual reporting} score is greater than $2$.
    \item \emph{Unreliable News Sources}: sources flagged as conspiracy-theory news producers or sources whose \emph{factual reporting} score is less than or equal to $2$.
\end{itemize}


\begin{table}[t]
    \caption{Summary of the used corpora. We see that more than half of the articles in NELA-GT-2020 are related to the topic of COVID-19. The labels for reliable, unreliable and partisan sources are obtained from Media Bias/Fact Check.}
    \label{tab:corpus-stats}
	\setlength{\tabcolsep}{3pt}
	\scriptsize
    \centering
    \begin{tabular}{lr}
        \multicolumn{2}{c}{\textbf{NELA-GT-2020}} \\
        \toprule
        Total Articles & \textasciitilde1.8M \\
        COVID-19 Articles & \textasciitilde1M \\ 
        Total Sources & 493 \\ 
        Labeled Sources & 316 \\ 
        Reliable Sources & 122 \\
        Unreliable Sources & 194 \\
        Partisan Sources & 162 \\
    \end{tabular}
    ~
    \begin{tabular}{lr}
        \multicolumn{2}{c}{\textbf{Scientific References}} \\
        \toprule
        COVID-19 Papers (CORD-19) & \textasciitilde300K \\ 
        Scientific Domains (SciLens) & \textasciitilde1K \\ 
        References in NELA-GT-2020 & \textasciitilde200K \\
    \end{tabular}    
\end{table}



\spara{Scientific References.}
We enhance the news collection described above, by extracting the external scientific references of news articles, i.e., the outgoing hyperlinks from the main body of the news articles.
We also extract the context of each reference, i.e., the passage of the news article that surrounds this reference.
We consider two repositories of references provided by \emph{CORD-19} and \emph{SciLens}.

One of the most prominent collection of papers related to \emph{COVID-19}, consisting of peer-reviewed papers as well as preprints and other historical coronavirus research, is \emph{CORD-19} \cite{DBLP:journals/corr/abs-2004-10706}.
We use the \emph{2021-06-14} release of \emph{CORD-19}, which contains a total of $310{,}833$ papers.
%
%
%

The second source of scientific references comes from \emph{SciLens} \cite{DBLP:conf/www/Smeros0A19}.
\emph{SciLens} provides a list of the top-1000 university domains (as indicated by \rurl{CWUR.org}), enhanced with a manually curated list of open-access publishers and grey literature databases.
Indeed, these scientific references are more prevalent in news than the \emph{CORD-19} papers, because their writing style and terminology used is typically more oriented towards a non-expert audience.


\section{Content Indicators}\label{sec:content-indicators}
%
%
In this section, we introduce two content-based indicators that we use to align news sources.
Particularly, we introduce an indicator regarding the shared content and an indicator regarding the semantic shift of terms between sources.

%


\begin{figure}
    \centering
    \includegraphics[width=.6\linewidth]{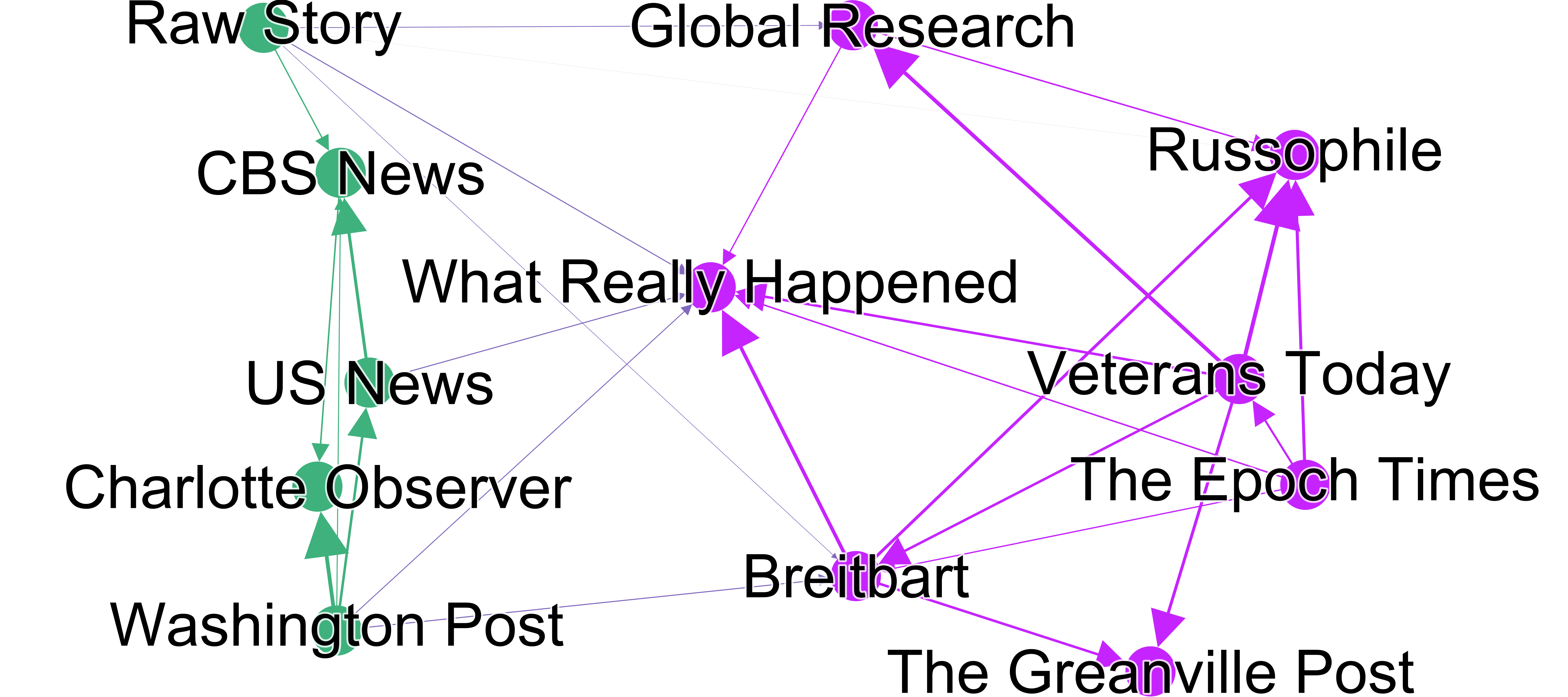}
    \caption{Example of a subgraph of the Content Sharing Network where nodes, representing sources, are connected by directed edges denoting the direction of the copied content between sources. Node color indicates the reliability class of the source (green for \emph{Reliable}, purple for \emph{Unreliable}), and edge width indicates the amount of content copied.
    }
    \label{fig:csn_example}
\end{figure}

\subsection{Copy Indicator}\label{sec:copy-indicator}
Content Sharing Network (CSN) is a model of content replication by sources in the news landscape. 
The sharing of news articles has been shown to be a common factor between news sources that adopt similar narratives around certain topics, which also correlates with the credibility of these sources \cite{DBLP:conf/icwsm/HorneNA19}.
Figure \ref{fig:csn_example} illustrates how sources are related in a CSN, where articles are copied from source to source.


The CSN is modeled as a directed graph where nodes represent news sources and edges indicate sources that copy articles verbatim from one another. 
Edges weights are proportional to the amount of content copied between the connected sources. 
The adjacency matrix $C$ of such network represents the affinity between the news sources.
We obtain this matrix using the method proposed by Horne et al. \shortcite{DBLP:conf/icwsm/HorneNA19} which consists of computing document vector representations for news articles using a TF-IDF bag-of-words representation. 
Articles are considered verbatim copies of each other if the cosine similarity between their vectors is greater than a threshold of $0.85$, and the direction of the copying is determined by the publication date of the article. The similarity threshold is defined following the recommendations from \citet{DBLP:conf/icwsm/HorneNA19}.

The final adjacency matrix is obtained by aggregating all copied articles at the source level. %
Thus, a directed edge from node $i$ to $j$ exists if source $j$ copies articles from source $i$.
%
The complement of the degree of relatedness distance between sources $i$ and $j$, is given as a function of the weight of the edge $(i,j)$ and is defined as:
\begin{align*}
    d_{cpy}(i,j) = 1 - \frac{|A_i \cap A_j|}{|A_j|}
\end{align*}
where $A_i$ and $A_j$ are articles published by sources $i$ and $j$; thus, their intersection should contain articles from source $i$ copied by source $j$.
The value of $d_{cpy}$ increases as fewer articles are copied from $i$ to $j$ and decreases as more articles in $j$ are copied from $i$.



\begin{table}[t]
    \caption{Semantic shift of the term ``antiviral''. We observe a contextual shift of the word. In the top two cases, the term is used to describe alternative medicine with herbs, while in the bottom two cases, the term is used with its ordinary (scientific) connotation.}
    \label{tab:shift-example}
	\setlength{\tabcolsep}{3pt}
	\scriptsize
    \centering
    \begin{tabular}{p{.3\linewidth}p{.6\linewidth}}
        \textbf{Source} & \textbf{Usage} \\
        \toprule
            Modern Alternative Mama &  \{...\} \textbf{these specific herbs have strong antiviral actions}, including against other strains of coronavirus. \\
            \midrule
            Healthy Holistic Living & \{...\} Garlic is known to have potent antibacterial, \textbf{antiviral}, antifungal and antiprotozoal abilities. \\
            \midrule
             The Guardian & \{...\} overwhelming emergency departments and causing governments to overspend on \textbf{antiviral medications}.\\
             \midrule
             The Washington Post & \{...\} although the \textbf{antiviral drug} remdesivir has been shown to help some patients \{...\}\\
        \bottomrule
    \end{tabular}    
\end{table}

\subsection{Shift Indicator}\label{sec:shift-indicator}
We analyze how specific technical terms are used differently between news sources. 
Different uses of a certain term in two pieces of text can occur if that same term is used in a different context in each of the texts. 
Semantic shift is the process through which the usage of a given word drifts when compared across different sources. Specifically, we consider the lexical semantic shift, which posits the semantics of a word to be defined by its contextual relationships to other lexicons \cite{cruse1986lexical}.
We argue that significant contextual shifts of topic-related words may serve as a signal of source disagreement, i.e., two sources using a certain target word in significantly different contexts may indicate that they use such words with different intents.
An illustrative example is shown in Table~\ref{tab:shift-example}.
Note, in both examples, the word antiviral is still used to indicate ``something that is effective against viruses''; however, the contexts give different connotations to what the antiviral product is.
 
Semantic shift has been used extensively in computational linguistics studies of language evolution \cite{DBLP:conf/emnlp/HamiltonLJ16} and, more recently, in studies quantifying the linguistic differences across domains \cite{DBLP:conf/nips/YinSP18, DBLP:conf/acl/SchlechtwegHTW19}.
In our method, we use semantic shift as an indicator of agreement among sources as it helps to uncover unique narratives created by unreliable sources, especially those based on conspiracy theories, deviating significantly from the narratives from reliable media.

The semantic shift between two sources $i$ and $j$ is measured by the deviation in the usage of words they have in common. 
Specifically, we define semantic shift as the aggregated distance between word embeddings for terms in the common vocabulary of sources $i$ and $j$. However, because the word embeddings are trained independently from each other, they cannot be directly compared. For example, suppose that $v_a$ and $v_b$ denote word vectors for the word \emph{virus} learned from the sources \textit{The Washington Post} and \textit{Global Research}, respectively. The cosine distance $d_{cos}(v_a, v_b)$ is not meaningful unless we first create a mapping between the embedding spaces of each source. This mapping can be achieved by applying an orthogonal transformation to one of the embedding spaces to minimize the sum of the pairwise Euclidean distances between word vectors of the common vocabulary. Being orthogonal means that this transformation preserves the inner product of the embeddings in the transformed space; for that reason, this mapping is also called embedding \emph{alignment} \cite{DBLP:conf/emnlp/HamiltonLJ16,joulin2018loss}.

Finding the best alignment of two embedding spaces is not a trivial task. 
Learning a transformation from all the words in the common vocabulary is often undesired, as the objective of the mapping is to minimize the distance between every pair of word vectors, hence minimizing the distance between words that are potentially semantically distinct \cite{DBLP:conf/nips/YinSP18}. To learn alignments between word embeddings, we employ the state-of-the-art self-supervised semantic shift (S4) method \cite{DBLP:conf/aaai/GruppiCA21}, which is designed to select the best words for generating a mapping between two embeddings.
This procedure is applied to embeddings trained using Word2Vec \cite{DBLP:journals/corr/abs-1301-3781}.


Once we train and align the embeddings, we compute the semantic distance between sources $i$ and $j$ as the average cosine distance between the top $10\%$ most frequent words in $i$ and $j$ (stop words excluded). Thus, the distance between sources $i$ and $j$ is defined as:
%
\begin{align*}
    d_{sem}(i,j) = \frac{\sum_{v \in V_i \cap V_j}{cos(emb_i(v), emb_j(v))}}{\left|V_i \cap V_j\right|}
\end{align*}
where $V_i$ and $V_j$ are the vocabularies of sources $i$ and $j$, $emb_i(v)$ and $emb_j(v)$ compute the embeddings representation of word $v$, and $cos$ computes the cosine distance between the embeddings. Additionally, $V_i$ and $V_j$ may be replaced with subsets of the common vocabulary to avoid using every word in the analysis (e.g., filter for the most frequent words).


\begin{table}[t]
    \caption{Usage of scientific jargon when citing a report by \emph{CDC} \cite{cdcmentalhealth2020}. We highlight that the citation context of \emph{TheNewYorker} is semantically closer to the report than the citation context of \emph{RedState}.}
    \label{tab:jargon-example}
	\setlength{\tabcolsep}{3pt}
	\scriptsize
    \centering
    \begin{tabular}{p{.2\linewidth}p{.75\linewidth}}
        \textbf{Source} & \textbf{Reference Context} \\
        \toprule
            TheNewYorker & In June, just three months into a historic health crisis, a survey by the Center for Disease Control and Prevention found that forty per cent of Americans were already struggling with at least one \textbf{mental-health issue}.\\
            \midrule
            RedState & It is no wonder that many Americans have \textbf{lost their faith} throughout 2020. Too many leaders have been inconsistent in their actions minus their continued breaches of the public trust.\\
        \bottomrule
            {} & {}\\
        \textbf{Reference} & \textbf{Title} \\
        \toprule
            CDC & \textbf{Mental Health}, Substance Use, and Suicidal Ideation During the COVID-19 Pandemic — United States, June 24–30, 2020.\\
        \bottomrule
    \end{tabular}    
\end{table}

\section{Reference Indicators}\label{sec:reference-indicators}
In this section, we introduce the reference indicators that we used to align news sources.
Particularly, we introduce two dedicated scientific indicators, namely, the usage of scientific jargon and the citation stance.
These indicators are \emph{reference indicators}, i.e., they define a distance among sources given a common (scientific) reference.

\subsection{Reference Context Extraction}
To compute the \emph{reference indicators}, we need the textual context of the references, i.e., the paragraph in which these references are cited.
To extract this context, we:
\begin{inparaenum}[i)]
    \item locate the references by parsing the raw HTML page of each news article of our data collection, and
    \item traverse the structural tree of the page to discover the most fine-grained text passage that contains the reference.
\end{inparaenum}
Currently, we do not support end-notes within articles, i.e., anchors at the bottom of articles where all the scientific references are listed, because it is a journalistic practice rarely appearing in our corpus.

\subsection{Jargon Indicator}

This indicator quantifies the scientific nature of the context in which a reference is used.
To estimate this indicator, we need a lexicon of terms ($jargon\_terms$ in the following) that are considered jargon in the scientific domain of our corpus.
Since, as we explain in \S\ref{sec:corpus}, our corpus contains news articles related to COVID-19, we use the vocabulary of \emph{CDC A-Z Index}\footnote{\url{https://www.cdc.gov/az}}, manually enhanced with common COVID-19 terminology.
%
%
After applying standard cleaning (e.g., punctuation removal), we compute the following distance:
\begin{align*}
d_{jar}(i,j) = |ctx_r(i) \; \cap \; ctx_r(j) \; \cap \; jargon\_terms|
\end{align*}
where $ctx_r(i)$ and $ctx_r(j)$ are the terms in the citation contexts of sources $i$ and $j$ for each common reference $r$.
%

We note that we do not aggregate for all common references between sources $i$ and $j$; hence, we do not limit to a single distance between these sources.
In this way, we encode the co-citation volume between sources $i$ and $j$, which is useful for our triplet sampling strategy (details in \ref{sec:triplet-sampling}).
After computing $d_{jar}(i,j)$, we apply \emph{Min-Max Normalization} in the interval $[0, 1]$ to comply with the previously-defined distances.
As we observe in Table~\ref{tab:jargon-example}, even such a simplistic metric is able to capture cases in which news sources completely distort the scientific message of the cited reference.
%

\begin{table}[t]
    \caption{Stance of news sources when citing (using the underlined hypertext) a webinar by \emph{CDC} \cite{cdccoviddeaths2020}. We highlight that \emph{The Truth About Cancer} uses more emotionally loaded words than \emph{FiveThirtyEight}.}
    \label{tab:stance-example}
	\setlength{\tabcolsep}{3pt}
	\scriptsize
    \centering
    \begin{tabular}{p{.2\linewidth}p{.75\linewidth}}
        \textbf{Source} & \textbf{Reference Context} \\
        \toprule
             FiveThirtyEight & \{...\} \underline{based on current CDC guidelines} \{...\} experts said that undercounting (deaths) was still more likely than overcounting.\\
        \midrule
            The Truth \newline About Cancer & Perhaps worst, the \textbf{CDC has continued to lie} about the death count by artificially inflating it. CDC \underline{guidelines} for determining COVID-19 deaths include: Anyone who tests positive, even if they died from other causes. Anyone who had COVID-19 symptoms, even if they aren’t tested. \\
        \bottomrule
            {} & {}\\
        \textbf{Reference} & \textbf{Title} \\
        \toprule
            CDC & Guidance for Certifying Deaths Due to Coronavirus Disease 2019 (COVID-19).\\
        \bottomrule        
        \end{tabular}
\end{table}

\subsection{Stance Indicator}

This indicator quantifies the sentiment charge of the context in which a reference is cited.
To measure this sentiment charge, we use the \emph{Multi-Genre Natural Language Inference} model \emph{BART} for zero-shot classification \cite{DBLP:conf/acl/LewisLGGMLSZ20}.
This model\footnote{\url{https://huggingface.co/facebook/bart-large-mnli}} computes the probability that we infer a certain \emph{hypothesis} given a \emph{premise}.
Thus, the model needs no explicit training on the downstream task of stance classification since the desired classes are provided implicitly in the \emph{hypothesis}.
After experimenting with various templates for \emph{premise} and \emph{hypothesis}, we report the ones that yield the most reliable results: 
\begin{align*}
\emph{premise} &= \emph{\textbf{reference context}} \\
\emph{hypothesis} &= \emph{``The stance of this example is \textbf{negative}''}
\end{align*}
The output of this model is a value in the interval $[0,1]$, denoting the probability a given premise implies our hypothesis.
We note that, by using this premise and hypothesis, we treat \emph{neutral} and \emph{positive} stances similarly, i.e., as \emph{non-negative} stances, because we want to highlight extremely negative stances (Table~\ref{tab:stance-example}).
Using this model we compute the following distance:
\begin{align*}
d_{ref}(i,j) = |stance(ctx_r(i)) \; - \; stance(ctx_r(j))|
\end{align*}
where $stance(.)$ computes the stance of the citation contexts of sources $i$ and $j$ for each common reference $r$.

Similarly as above, after computing $d_{ref}(i,j)$, we apply \emph{Min-Max Normalization} in the interval $[0, 1]$.
%
%
As we observe in Table~\ref{tab:stance-example}, this indicator distinguishes between the sentiment of sources towards a common reference.

\section{Unsupervised Source Embeddings}\label{sec:methods-ensemble}
The previous section described the heterogeneous indicators that we extract from each news source.
In this section, we describe how we combine these indicators in a unified framework to learn unsupervised representations of news sources.
The triplet loss function aims at coupling different parts of the input spaces (here, our indicators) into a single representation \cite{weinberger2009distance}.
The triplets sampling and embeddings training methods employed in this framework are well-established methods \cite{DBLP:conf/simbad/HofferA15} used mainly in learning-to-rank recommendation systems \cite{DBLP:conf/sigir/ChenQZX16,DBLP:conf/www/WangSCJLHC21}.

\subsection{Triplets Sampling}\label{sec:triplet-sampling}
%
%
Our goal is, using the distances defined by the indicators, to discover pairs of similar sources and pairs of dissimilar sources.
%
%
%
%
By joining these two sets of pairs, we create triplets of the form (\emph{anchor, positive, negative}), where \emph{anchor} is the common element of the pairs, \emph{positive} is the element similar to the \emph{anchor}, and \emph{negative} is the element dissimilar to the \emph{anchor}.
For simplicity, in the following, we will refer to these triplets as (\emph{a, p, n}).

We note that these triplets may not occur from the same indicator, i.e., the positive pair may occur from an indicator that is more appropriate for capturing the affinity between sources, and the negative pair may occur from an indicator that is more appropriate for capturing the disparity between sources.
In our experimental evaluation (\S\ref{sec:experiments-indicators}), we evaluate each indicator in its ability to produce good positive and negative pairs as well as full triplets.



\subsubsection{Positive Pair Sampling.}
We use the distances computed for each indicator to generate pairs of similar sources.
For all indicators we introduce in \S\ref{sec:content-indicators} \& \S\ref{sec:reference-indicators}, short distance denotes similarity.
Given an indicator $f$ ($copy, shift, jargon,$ or $reference$), we generate a positive pair of similar sources $i, j$ with a probability inversely proportional to the distance between $i$ and $j$:
\begin{align*}
    pp_f(i,j) = \frac{d_f^{-1}(i,j)}{\sum_{k}d^{-1}_f(i,k)} & ~ \forall j \neq i
\end{align*}
We draw $l$ positives samples from this distribution for each indicator and each source in the dataset, producing a total of $l$ positive source pairs $(a, p)$.

\subsubsection{Negative Pair Sampling.}
For negative sampling, we employ two strategies.
For some indicators (e.g., the stance indicator), a large distance between sources denotes opposing sentiment, thus disagreement (e.g., the sources in Table~\ref{tab:stance-example}).
Hence, we use the inverse distribution we used for generating positive pairs to generate negative pairs:
\begin{align*}
    np_f(i,j) = 1 - pp_f(i,j) & ~ \forall j \neq i
\end{align*}
Similarly as above, we draw $l$ negative samples from this distribution for each indicator and each source in the dataset, producing a total of $l$ negative source pairs $(a, n)$.

Nonetheless, there are indicators (e.g., the copy indicator) for which a large distance between sources does not necessarily denote disagreement; it only denotes the absence of agreement.
In these cases, we draw the negative pairs uniformly from the set of sources.
%

Finally, we employ a cleaning heuristic to increase the accuracy of our triplets (detailed experiment in \S\ref{sec:experiments-indicators}).
Specifically, we make sure that we do not select a negative pair $(a, n)$ which we have already selected as positive pair $(a, p)$:
\begin{align*}
    (a, p) \wedge (a, n) \Rightarrow p \neq n
\end{align*}

\subsection{Embeddings Training}\label{sec:training-embeddings}

Once we extract all the triplets, we use them for training a dense representation model for news sources with the \emph{Triplet Margin Loss} \cite{DBLP:conf/bmvc/BalntasRPM16}.
The learning objective of \emph{Triplet Margin Loss} is to minimize the distance between an anchor and a positive sample while maximizing the distance between the anchor and the negative sample.

The procedure we employ is the following.
First, we initialize the embeddings for all the sources into a low-dimensional, dense vector space by randomly setting the weights in the embedding layer following a normal distribution $\mathbb{N}(0,1)$.
%
%
%
Then, given the input triplets $(a, p, n)$, we train these embeddings by minimizing the loss function $L$:
\begin{align*}
    L(a, p, n) = max\{d(a, p) - d(a, n) + M, 0\}
\end{align*}
where 
$d$ is the distance function, and $M$ is the margin parameter that controls the gap between positive and negative distances. The larger $M$ is, the larger is the gap between $d(a,p)$ and $d(a,n)$.
We train the embeddings over several epochs until convergence and then use them as the representation of the news sources.
%

The parameters of this method are the margin $M$, the distance function $d$, and the size of the output vectors $s$.
We release the optimal training parameters as well as the trained sources embeddings in our code release (\S\ref{sec:conclusions}).

%

\section{Experiments}\label{sec:experiments}
Our experimental evaluation is three-fold; first, we evaluate the indicators individually, then we evaluate the source embeddings on the downstream task of source reliability classification, and finally, we perform an unsupervised clustering where we analyze the patterns in the news sources captured by the learned features.
In the following experiments, the labels from Media Bias/Fact Check are used, and word embeddings for the semantic shift are trained using Word2Vec with dimension $100$, context window of $10$, and minimum word count of $20$.
The parameters for SciLander are margin $M=1$, vector size $s=50$, and distance $d$ used in the loss function is the cosine distance.

\subsection{Indicator Coverage}\label{sec:experiments-indicators}

In our first experiment, we measure the overlap of the introduced indicators in terms of source and triplet coverage.
We also measure the accuracy of the triplets computed by these indicators.

\begin{figure}[t]
    \setlength{\tabcolsep}{3pt}
	\tiny
    \centering
    \includegraphics[width=.3\linewidth]{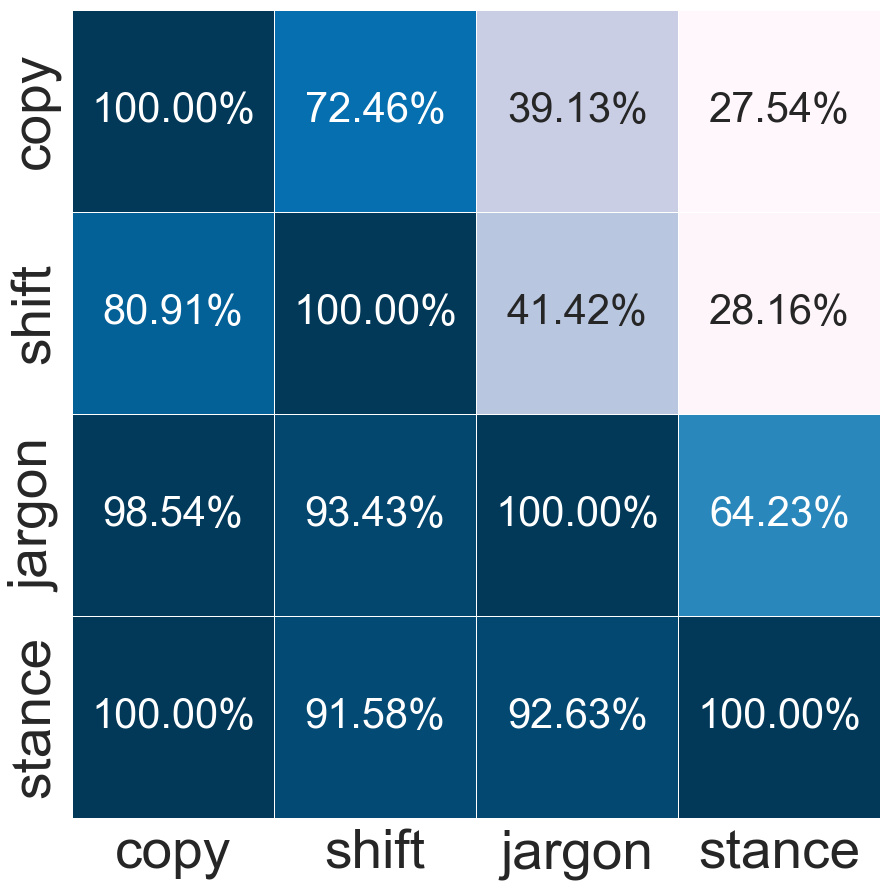}
    \includegraphics[width=.3\linewidth]{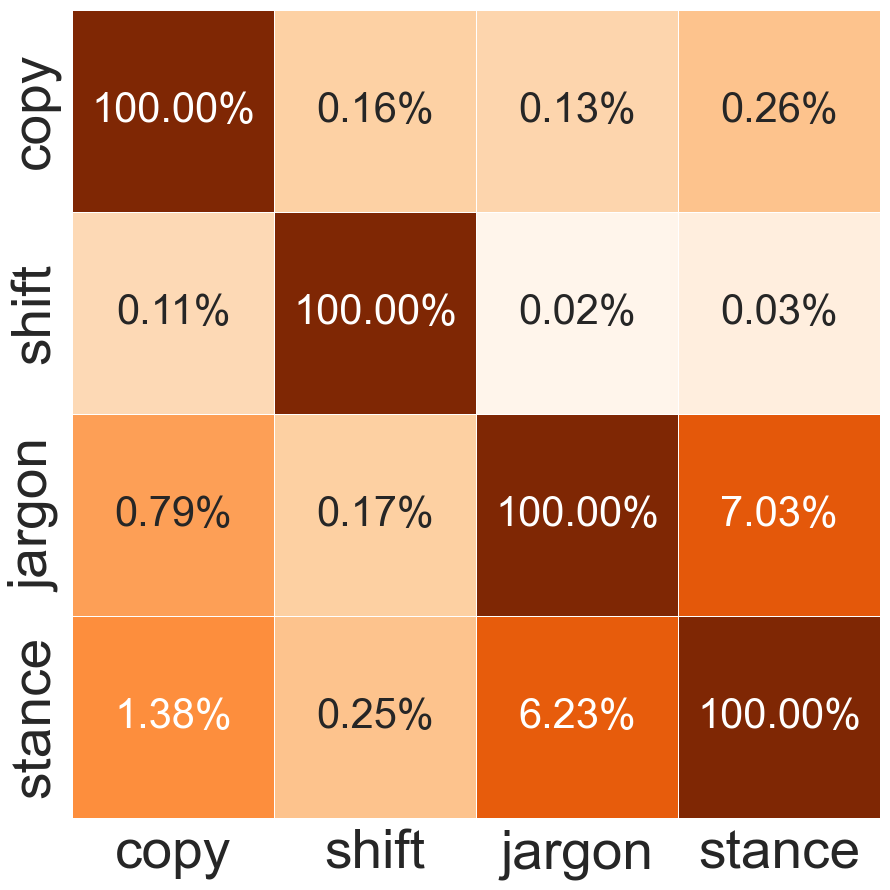}
    \begin{tabular}{lrrrr}
    {} \\
    \textbf{Indicator} & ${\mathbf{AUROC_p}}$ & ${\mathbf{AUROC_n}}$ & ${\mathbf{AUROC_f}}$ & \textbf{\#sources} \\
    \toprule
        copy & $72.7\%$ & $51.0\%$ & $36.3\%$ & $257$\\
        shift & $61.9\%$ & $60.8\%$ & $41.8\%$ & $308$\\
        stance & $89.7\%$ & $73.3\%$ & $68.3\%$ & $87$\\
        jargon & $81.9\%$ & $51.0\%$ & $42.9\%$ & $126$\\
        \midrule
        overall & $77.0\%$ & $69.7\%$ & $57.5\%$ & $316$\\
        \bottomrule
        \end{tabular}
    \caption{Overlap of indicators in terms of source coverage (top left) and triplet coverage (top right); AUROC of the positive part, negative part, and full triplets (bottom). Although the sources covered by most indicators heavily overlap, their triplets are quite unique. Also, there is a trade-off between the source coverage of the indicators and their AUROC.}
    \label{fig:indicators_coverage}
\end{figure}

We define the source coverage ($sc$) and the triplet coverage ($tc$) between two indicators $i, j$ as follows:
{\small
    \begin{align*}
        sc(i,j) = \frac{|src(i) \cap src(j)|}{|src(i)|}, \;
        tc_(i,j) = \frac{|trpl(i) \cap trpl(j)|}{|trpl(i)|}
    \end{align*}
}%
where $src(.)$ and $trpl(.)$ compute the distinct set of sources and triplets covered by a given indicator.
We note that $sc$ and $tc$ are non-symmetric; consequently, the heatmaps in Figure~\ref{fig:indicators_coverage} are also non-symmetric. 

To measure the accuracy of the computed triplets, we use the metric \emph{Area Under the Receiver Operating Characteristics} (AUROC), which measures the \emph{True Positive Rate} over the \emph{False Positive Rate}.
We also break down the AUROC of the triplets into 
\begin{inparaenum}[i)]
\item the $AUROC_p$ of the positive part of the triplets $(a, p)$,
\item the $AUROC_n$ of the negative part of the triplets $(a, n)$, and
\item the $AUROC_f$ of the full triplets $(a, p, n)$.
\end{inparaenum}
Specifically, for each individual AUROC, we consider the following as true positives:
{\small
    \begin{align*}
    AUROC_p &: \{(a, p) \; s.t. \; label(a) = label(p)\} \\
    AUROC_n &: \{(a, n) \; s.t. \; label(a) \neq label(n)\} \\
    AUROC_f &: \{(a, p, n) \; s.t. \; label(a) = label(p) \\ 
            & \hspace{0.1\textwidth} \wedge label(a) \neq label(n)\}
    \end{align*}

}%


As we observe in Figure~\ref{fig:indicators_coverage}, although the sources covered by some indicators heavily overlap, the contributed triplets are quite unique.
Indicatively, the stance indicator covers $27.5\%$ of the sources, totally overlapping with the copy indicator.
However, the contributed triplets of the stance indicator are different from the contributed triplets of all the other indicators and also more accurate.
Indeed, we see that there is a trade-off between the source coverage of the indicators and the AUROC.
Hence, the more specific the indicator is (e.g., the stance indicator), the better AUROC it has.

Finally, we observe that the overall AUROC for positive and negative pairs (AUROC$_p$ and AUROC$_n$, respectively) are above the 50\% baseline of a random positive (or negative) pair selection is truly positive (or negative).

It should be noted that the AUROC for complete triplets (AUROC$_f$) is lower than 50\%.
This happens because the choice of the final triplets involves two independent decisions: the choice of the positive sample, and the choice of the negative sample. As noted above, each choice has a chance of success of 50\% if chosen at random. Thus, for a triplet to be correctly selected, the random baseline is that a correct positive pair is chosen \emph{and} a correct negative pair is chosen, which results in a $0.5 \times 0.5 = 0.25$, or 25\% baseline chance.
%
%
%
%
%
As we see in the following experiments, the model for training source embedding is robust to noisy triplets as it yields highly accurate results in all the downstream tasks we use it.

\subsection{Offline Source Classification}\label{sec:offline-classification}

In this experiment, we evaluate the computed embeddings on a downstream classification task.
We assume that, for all sources in our corpus, we have (offline) access to a significant fraction of their history of published articles.
%

\subsubsection{Baselines.}
For this task, we implement baselines using \emph{Stylistic Text Features}, \emph{Contextualized Embeddings}, and \emph{Co-citation Embeddings}, as well as combinations of the above.

\emph{Stylistic Text Features.} We utilize stylistic text features from Horne et al. \cite{DBLP:journals/corr/HorneA17} 
aggregated at the source level as representations.
These features include, among others, the number of:
\begin{inparaenum}[i)]
part of speech tags,
punctuation symbols, and
capitalized words,
\end{inparaenum}
which are the features that are typically used in news classifiers.

\emph{Contextualized Embeddings.} We compute BERT \cite{DBLP:conf/naacl/DevlinCLT19} embeddings for a total of $32$ tokens from the title and the opening paragraph of the article, and average them for each source.
Similarly, we compute SciBERT \cite{DBLP:conf/emnlp/BeltagyLC19} instead of BERT embeddings, which have been shown to lead to better performance in tasks involving scientific text.
The configuration parameters of both BERT and SciBERT are those suggested in a widely used release of this model \cite{DBLP:journals/corr/abs-1910-03771}.

\emph{Co-Citation Embeddings.}
We compute a co-citation graph of sources based on their scientific references.
We weight this graph either uniformly for each common reference, or by emphasizing the uniquely used references, using their TF-IDF score.
In the overall graph, we run node2vec \cite{DBLP:conf/kdd/GroverL16} to extract source embeddings.

\emph{Joint Embeddings.}
The \emph{Contextualized Embeddings} and the \emph{Co-Citation Embeddings} capture two different modalities of news sources; their content and citation behavior.
Thus, we create a joint representation 
by concatenating the two embeddings.
Since the dimensionality of the joint embeddings is high,
we apply Principal Component Analysis to reduce it and compare it with other baseline representations.

\subsubsection{Evaluation.}
\begin{figure}[t]
    \centering
    \includegraphics[width=.85\linewidth]{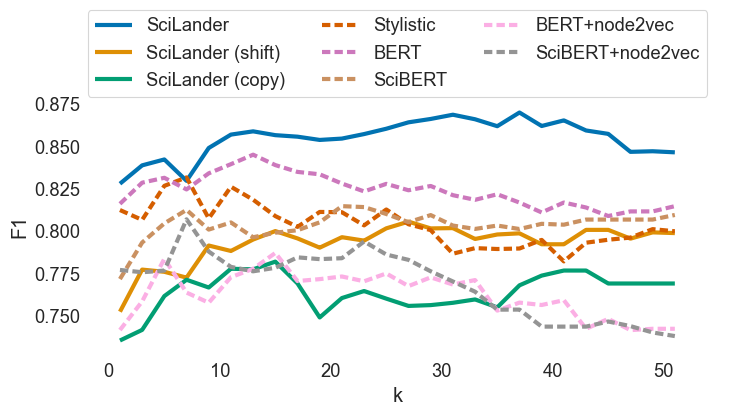}
    \caption{F1 scores using k-nearest neighbors classifiers over the source embeddings representations computed by SciLander and the various baselines described in \S\ref{sec:offline-classification}. SciLander obtains the best F1 score (\emph{87\%}) for \emph{k=37}.}
    \label{fig:model_comparison}
\end{figure}

We test the usefulness of the learned representations in the problem of source veracity classification.
%
%
We use the embeddings computed by 
\begin{inparaenum}[i)]
\item SciLander trained on all indicators,
\item SciLander trained only on content indicators (shift or copy), and
\item the aforementioned baseline models,
\end{inparaenum}
  to train a Nearest Neighbors classifier in a 10-fold cross-validation setting.
Figure \ref{fig:model_comparison} shows the F1 score of each model for increasing values of k.

Relying uniquely on textual features limits classifiers to a restricted set of signals.
Our framework combines stylistic, semantic, and behavioral indicators to produce a representation that improves the separation of reliable and unreliable sources.
Thus, compared to traditional baselines such as stylistic features or features extracted by BERT, our embeddings show significant performance improvement.
Our method obtains the best F1 score ($\mathbf{87\%}$) for $\mathbf{k=37}$.

\subsection{Online Source Classification}\label{sec:online-classification}

In this experiment, we assume that we have two types of sources: 
\begin{inparaenum}[i)]
\item offline (known) sources, for which we have access to a significant fraction of their publication history, and
\item  online (newcomer) sources, for which we have access to a limited fraction of their publication history.
\end{inparaenum}
As assessing articles from newcomer sources might be a time-consuming task, we inspect the lowest fraction of articles that is needed to accurately classify these sources. 

The procedure that we employ is the following:
\begin{inparaenum}[i)]
\item we train embeddings for the \emph{offline sources} (as we explain in \S\ref{sec:training-embeddings});
\item we freeze these embeddings for the \emph{offline sources};
\item we train embeddings for \emph{online sources}, in the already shaped by the \emph{offline sources} embeddings space.
\end{inparaenum}

\begin{figure}[t]
    \centering
    \includegraphics[width=.85\linewidth]{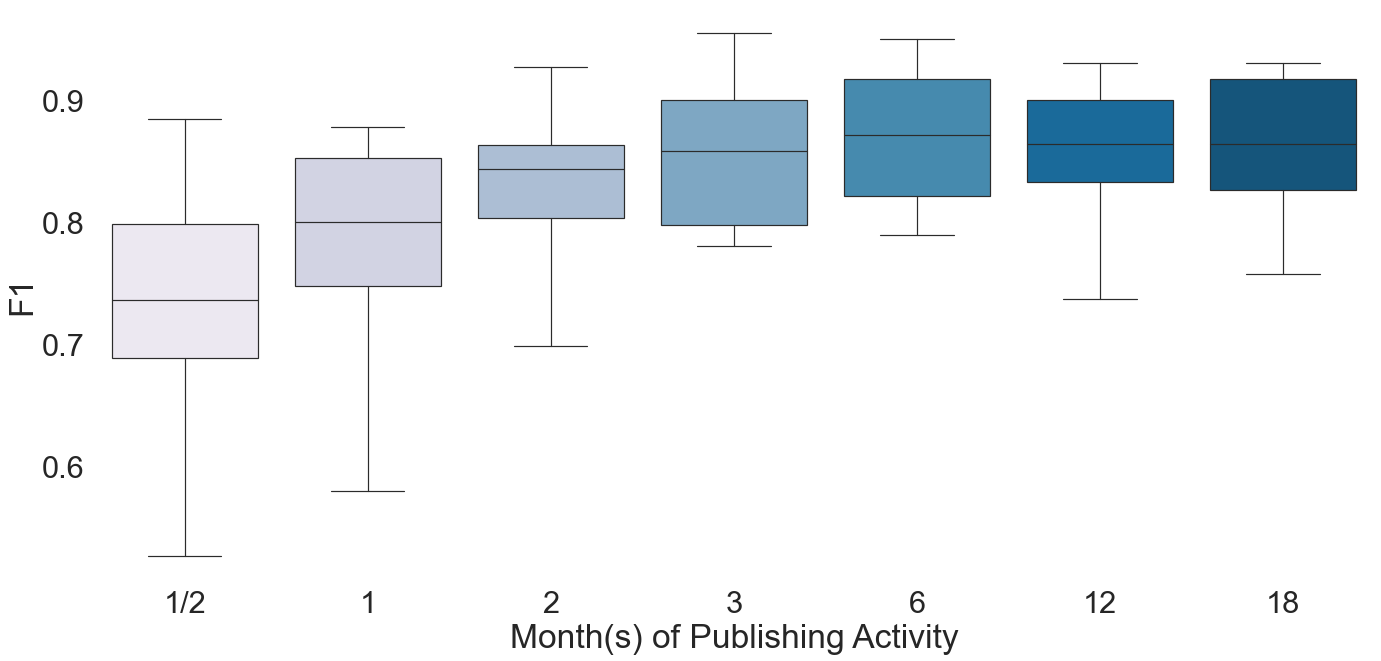}
    \caption{Learning curve (F1 score) for increasing fractions of articles from newcomer sources.
    SciLander is able to reliably (\emph{F1>85\%}) classify sources using only \emph{3} months of their publishing activity.}
    \label{fig:online_classification}
\end{figure}

We conduct the experiment on a 10-fold cross-validation setting.
In Figure~\ref{fig:online_classification}, we report the learning curve (F1 score) for increasing fractions of articles from newcomer sources in the same classification task described in \S\ref{sec:offline-classification}.
We note that the temporal axis is not in chronological order but sampled randomly from the entire corpus (e.g., we sample articles representing a 3-month publishing activity of an online source from the entire publishing activity of that source).
In that way, each temporal interval is independent of external events (e.g., the development of the vaccines), which affects the activity of most sources.
As we observe in Figure~\ref{fig:online_classification}, SciLander is able to reliably ($\mathbf{F1{>}85\%}$) classify sources, using only three months of their publishing activity.

\subsection{Source Clustering Analysis}

\begin{figure}[t]
    \centering
    \includegraphics[width=.25\textwidth]{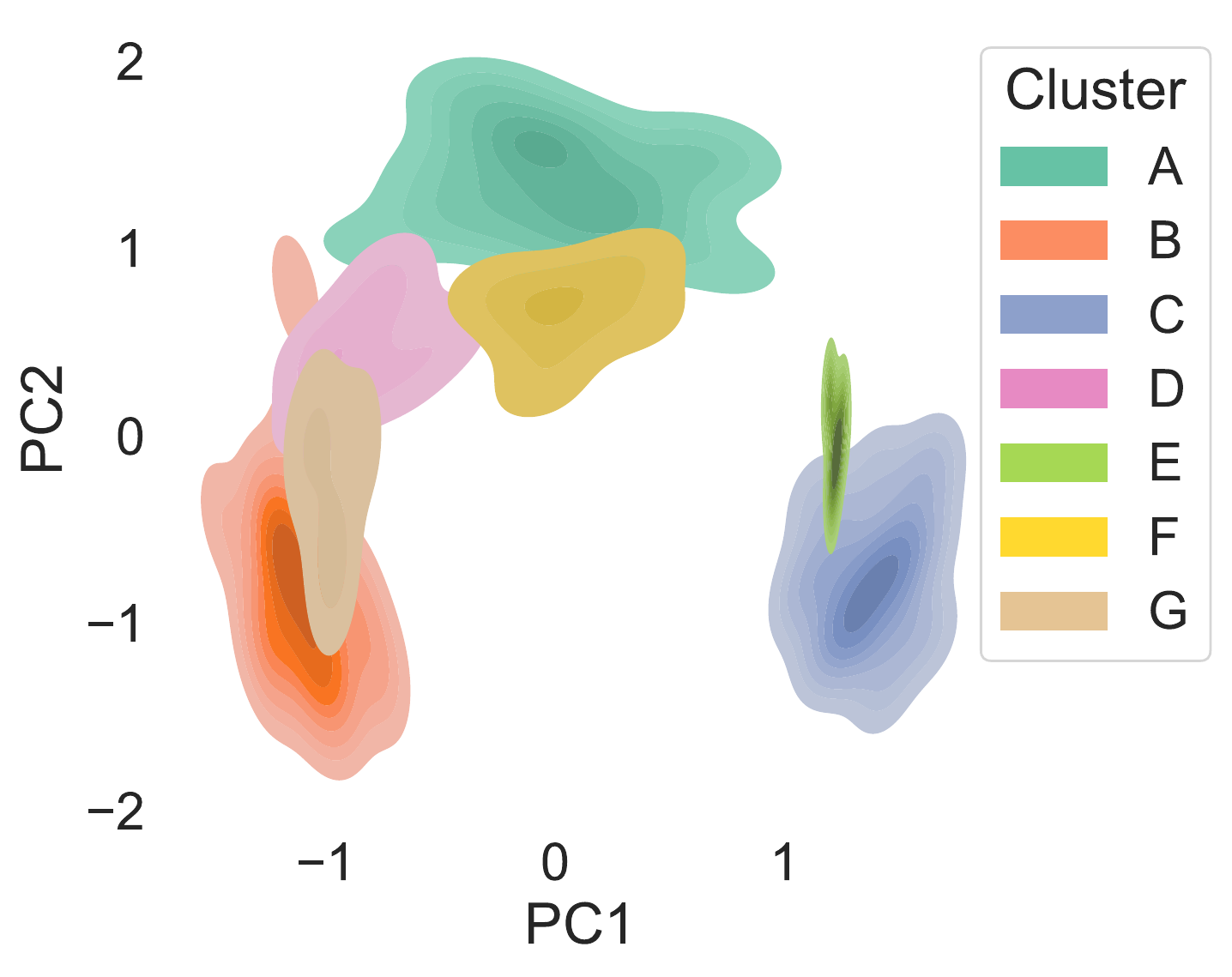}
    \caption{Kernel Density Estimation of the clusters}
    \label{fig:clusters}
\end{figure}

We conduct an unsupervised clustering experiment to investigate potential trends revealed by the features learned by SciLander.
Using the same embeddings from the previous experiments ($50$ dimensions, $M=1$), we apply DBSCAN clustering to the source vectors with the cosine distance as distance metric, minimum distance parameter $\epsilon=0.1$ and minimum cluster size $n=1$.
The resulting clusters are shown in Figure \ref{fig:clusters}; each of the 7 clusters is shown in different color shades and labeled from $A$ to $G$. 

\begin{figure*}[t]
    \centering
    \begin{subfigure}{0.25\textwidth}
    \includegraphics[width=\textwidth]{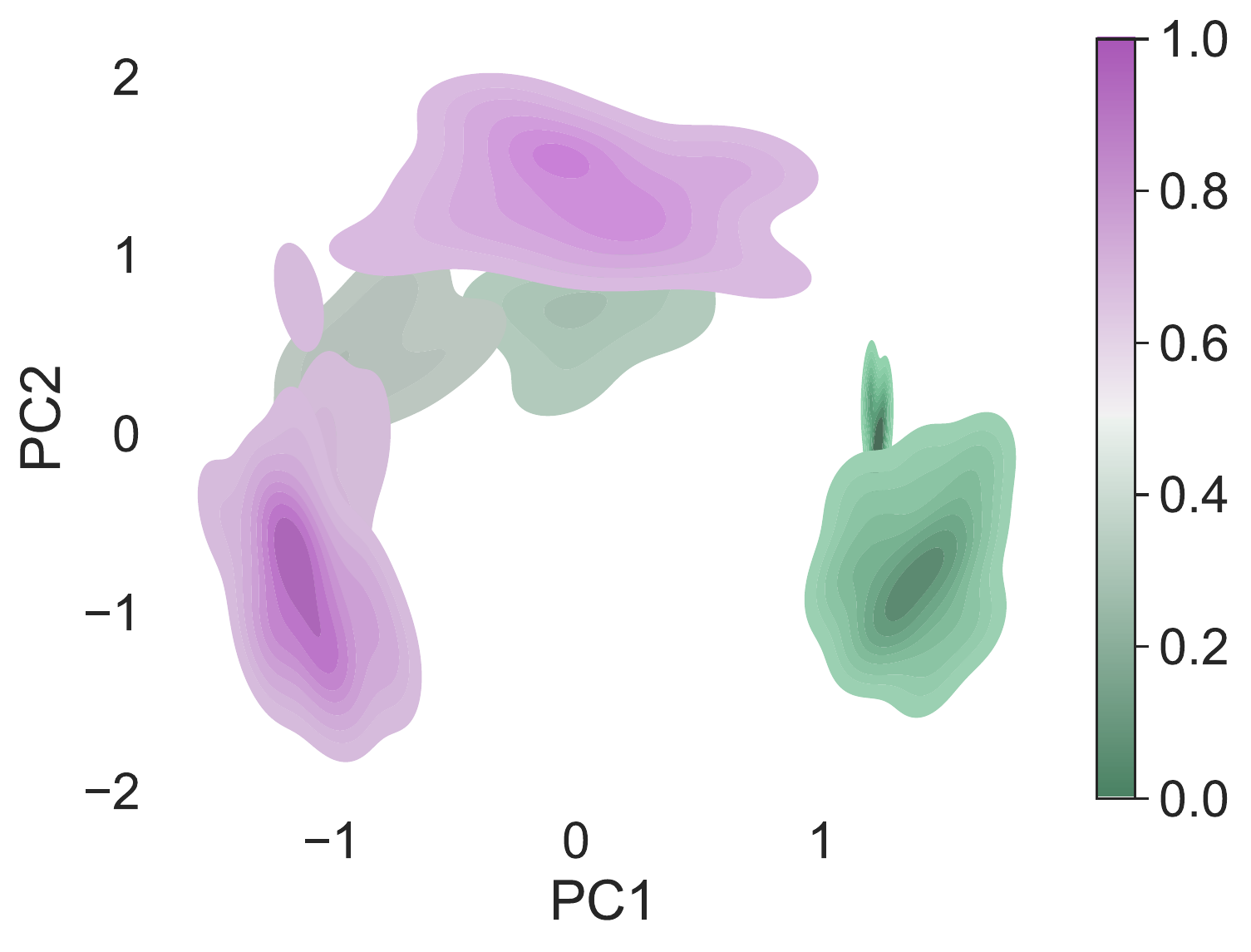}
    \caption{Density of Unreliable Sources}
    \label{fig:unreliable-density}
    \end{subfigure}
    \qquad
    \begin{subfigure}{0.25\textwidth}
    \includegraphics[width=\textwidth]{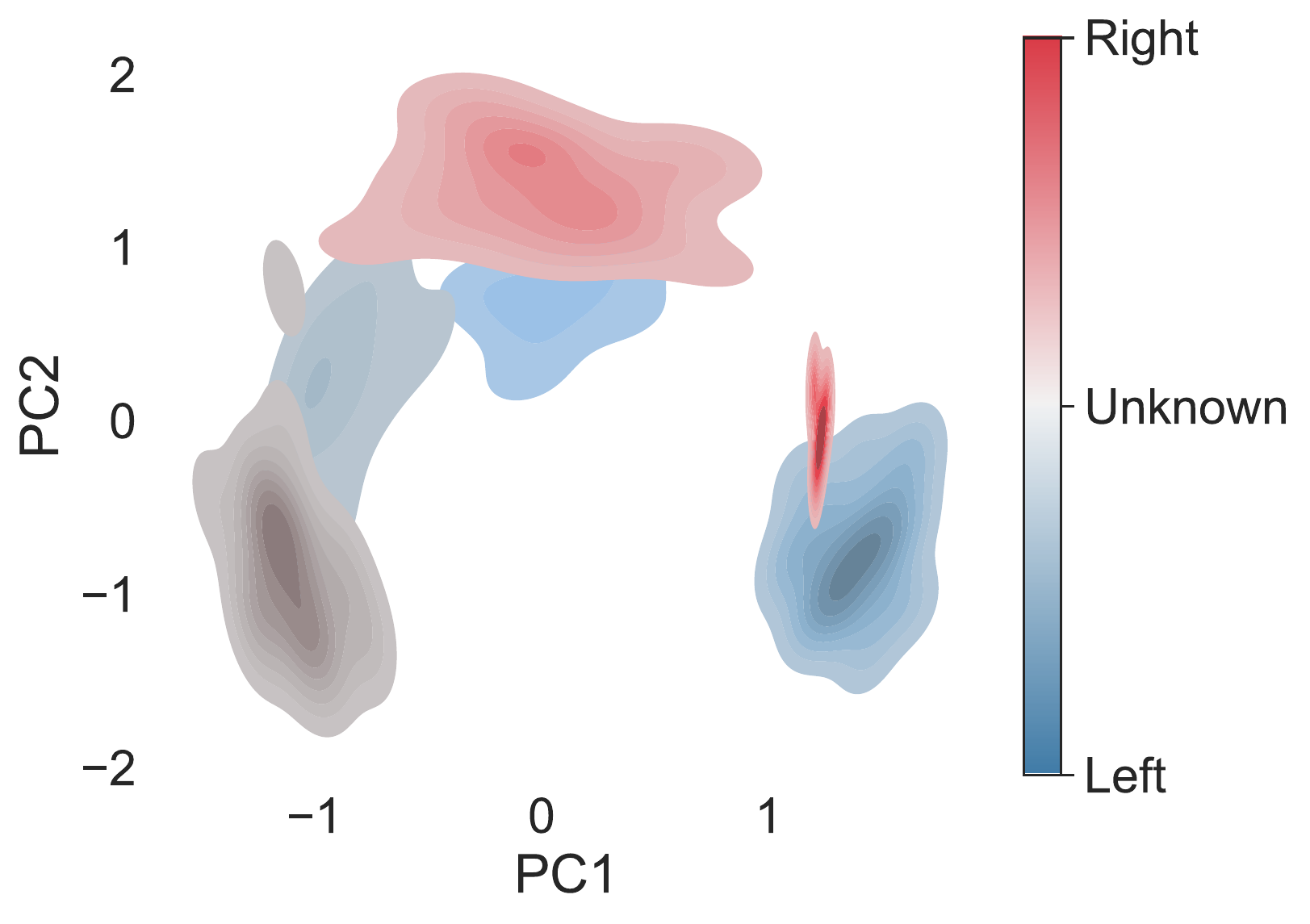}
    \caption{Density of Political Leaning}
    \label{fig:political-leaning}
    \end{subfigure}
    \qquad
    \begin{subfigure}{0.25\textwidth}
    \includegraphics[width=\textwidth]{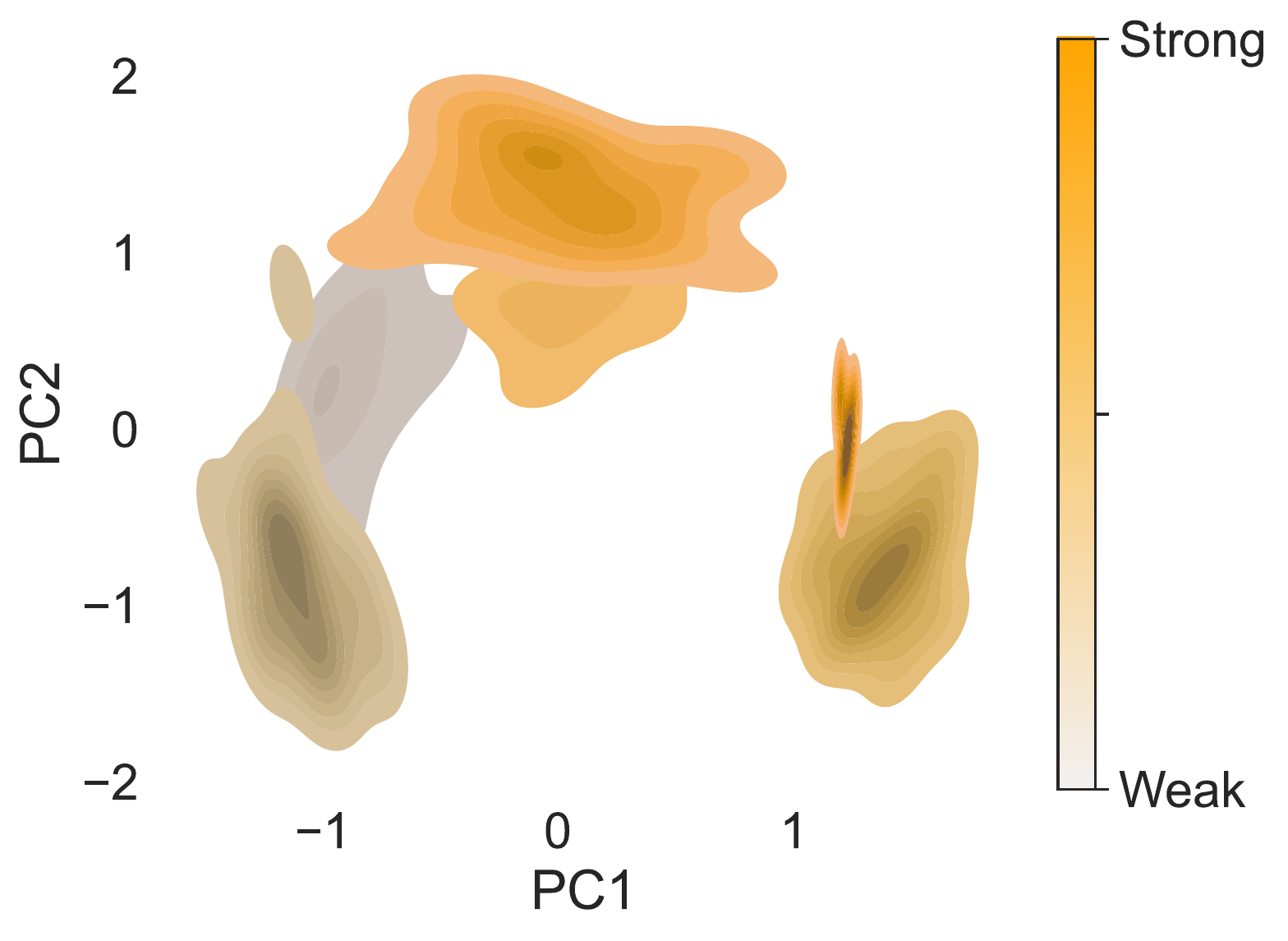}
    \caption{Density of Partisanship Bias}
    \label{fig:partisanship}
    \end{subfigure}
    \caption{Density analysis of the clusters computed by SciLander. Components PC1 and PC2, obtained from Principal Component Analysis (PCA) on the source embeddings, are the components with the highest explained variance ratio.}
    \label{fig:cluster_analysis}
\end{figure*}

We characterize the clusters quantitatively with respect to the density of unreliable sources, political leaning, and the level of partisanship bias aggregated across the news sources within them. 
For each cluster, we compute the proportion of unreliable sources to the total number of sources in the cluster. 
Figure~\ref{fig:unreliable-density} shows the density of unreliable sources within each cluster. 
This result suggests that the source embeddings carry information about source credibility when grouping them, even though credibility labels or related features were unknown to the model during training.

Clusters $C$ and $E$ contain no unreliable sources and hold mostly mainstream news sources such as The Washington Post, Vox, National Public Radio (NPR), and the Chicago Tribune. 
The clusters containing the largest proportions of unreliable sources are the clusters $A$, $B$, and $G$, and most sources in these clusters are websites that propagate conspiracy theories and promote pseudoscience. Details on the discovered clusters are shown in Table \ref{tab:clusters}.

These results show that the SciLander embeddings are able to group sources based on similar reliability.
Multiple clusters of relatively high purity with respect to reliability are created, some reliable (75\%-100\% reliable sources), some unreliable (0\%-30\% reliable sources).

We compute the overall political leaning of a cluster by averaging the political leaning scores of the sources within that cluster. 
Partisanship bias is obtained by the absolute value of leaning, scaled to a value in $[0,1]$, with $0$ indicating that there is no partisanship bias in the cluster, and $1$ indicating the maximum partisanship bias, where all sources in the cluster exhibit a strong political leaning.
The partisanship bias describes the agreement between the political leanings of sources within the cluster, and the magnitude of such leanings. The distribution of political leanings and partisanship bias are shown in Figures \ref{fig:political-leaning} and \ref{fig:partisanship}.
There is a noticeable disparity between the partisanship bias found in the two biggest unreliable clusters $A$ and $B$. Sources in cluster $A$ exhibit a strong bias, which is nearly absent in cluster $B$. We explore the particularities of these clusters next.

\begin{table}[t]
    \captionof{table}{\textbf{(U)}nreliability score (proportion of unreliable sources), average \textbf{(P)}artisanship bias score,  and core sources (nearest neighbors to the centroid) of the identified clusters.}
    \label{tab:clusters}  
    \setlength{\tabcolsep}{2.5pt}
    \centering
    \small
    \begin{tabular}{crrp{.77\linewidth}}
    \textbf{Cl.} & \textbf{(U)}    &   \textbf{(P)}   &   \textbf{Core Sources} \\
    \toprule
    \textbf{A}       &       .70    &       .25    &       NewsWars, Veterans Today, The D.C. Clothesline \\
\textbf{B}       &       .84    &       .03    &       Mercola, Healthy Holistic Living, Vaccine Reaction \\
\textbf{C}       &       .00    &       .11    &       The Washington Post, Vox, NPR \\
\textbf{D}       &       .25    &       .00    &       The American Conservative, Roll Call \\
\textbf{E}       &       .00    &       .20    &       Chicago Tribune \\
\textbf{F}       &       .12    &       .03    &       Washington Monthly, FiveThirtyEight, Atlantic \\
\textbf{G}       &       .80    &       .00    &       Ice Age Now \\

    \end{tabular}
\end{table}

\subsection{Different Types of Conspiracy Theories} \label{sec:different-conspiracies}

We observe two clusters with high density of unreliable sources (clusters $A$ and $B$). Both clusters include many unreliable news sources, and there exist qualitative differences between them, which we describe in this section.

To uncover qualitative differences between sources in clusters $A$ and $B$, we measure the shift in context between these clusters and the mainstream cluster $C$. 
Specifically, we computed the semantic shift across clusters of sources by training Word2Vec models $E_A$, $E_B$, and $E_C$ using articles from the core sources of each cluster and using the same hyper-parameters as in the previous experiments. 
Then, we extract the words with the highest cosine distance between pairs $(E_C,E_A)$ and $(E_C,E_B)$ to find the terms that most contribute to the deviation in the news from sources in $C$ to each of the unreliable clusters $A$ and $B$.

Let $S_A$ and $S_B$ be the lists of the $100$ words most shifted to $C$, from $A$ and $B$, respectively. 
We find that there is only one word in common between the $S_A$ and $S_B$: ``natural''.
To characterize the words in both lists, we identify words that refer
to people, entities and places, political issues, and health and nutrition. 
Examples of these words are given below and listed on Table~\ref{tab:shifted_words}.

\begin{table}[t]
    \caption{List of words from clusters \emph{A} and \emph{B} that are most shifted from the mainstream cluster \emph{C}. People and Places, and Political Terms appear as the most shifted words in cluster \emph{A}, suggesting that its sources push politically-oriented misinformation, while sources in cluster \emph{B} focus more on alternative health solutions.}
    \label{tab:shifted_words}
    \setlength{\tabcolsep}{3pt}
    \centering
    \footnotesize
    \begin{tabular}{p{0.29\linewidth}p{0.4\linewidth}p{0.21\linewidth}}
        \textbf{People and Places} & \textbf{Political Terms} & \textbf{Health} \\ \toprule
         Kamala Harris & BLM (Black Lives Matter) & Coronavirus  \\
         Bernie Sanders & Patriot & Food  \\
         Nancy Pelosi & Voting & Vaccines \\
         Mike Pence & Abortion & Doctors \\
         Alex Jones & Partisan & Mask \\ \bottomrule
    \end{tabular}
\end{table}

The largest group of words shifted in cluster A are related to individuals, entities, places (25\%), and political topics (12\%). 
Almost all individuals found are political figures (with a few exceptions). 
There are only 1.5\% of terms related to health and nutrition.
Many of these news outlets are conspiracy theory websites such as NewsWars, Veterans Today, and InfoWars. 
According to a Media Bias/Fact Check analysis\footnote{\url{https://mediabiasfactcheck.com/veterans-today}}, these sites often publish hate-speech-filled content in addition to misleading or false information.

In contrast, the largest group of shifted words was detected in cluster B (21.5\%), with only 2\% people and 1\% related to political
topics. 
According to MBFC journalists\footnote{\url{https://mediabiasfactcheck.com/mercola}}, these sources promote alternative health notions, sell questionable products and supplements, and promote antivaccination positions with pseudoscience-based arguments. 

Based on this, we conclude that while cluster A is a cluster of mostly politically-unreliable news sources covering COVID-19 stories mixed with other political topics, cluster B is much more focused on covering alternative medicine-based misinformation with slight political leaning, presumably to appeal to individuals with different political opinions.
On these sites, health-based information is often mixed with promotion and affiliate links to sites selling alternative medicine products and supplements.
Our method is able to properly distinguish these different types of COVID-19 misinformation, without explicitly training on related features.




\section{Discussion}\label{sec:discussion}

\spara{SciLander is a method for embedding news sources.}
The results of the experiments (\S\ref{sec:experiments}) show that the representations learned from SciLander outperform other state-of-the-art feature models.
Despite the final representation being a set of autoencoded features (i.e., embeddings learned from a neural network), it is directly explained by the product of a combination of the aforementioned indicators.

The applications of these learned features are not restricted to classification tasks.
They can be used in any scenario where similarities between news sources are needed, such as in clustering analysis (details in \S\ref{sec:different-conspiracies}) and recommendation systems.

SciLander, like most other AI/ML methods, is heavily data-driven. 
It uses signals found in the text of news articles to infer the relationships between sources.
The above can cause SciLander to make biased decisions, especially if the input data is biased towards/against societal groups, such as underrepresented minorities and other vulnerable groups.
We argue that SciLander, when deciding what content to recommend or promote, can provide assistance in human decision-making but not replace human judgment.

\spara{The indicators used by SciLander complement each other.}
The experiments shown in \S\ref{sec:experiments} demonstrate that the embeddings performed better in classification tasks when all four indicators are combined (copy, shift, jargon, and stance).
The latter suggests that the indicators worked in a complementary manner, where a mistake made by one indicator is corrected by the other indicators.
Furthermore, as seen in Figure \ref{fig:indicators_coverage}, some indicators were better suited to detect negative pairs. 
For example, the indicator \emph{copy} had a $AUROC$ score of 72.7\% for positive samples and 51\% for negative samples. 
This result can be explained by the fact that while the presence of copy behavior between sources $x$ and $y$ is an indicator of similarity between $x$ and $y$, its absence does not necessarily imply that $x$ and $y$ are very distinct.
In short, source $x$ not copying from $y$ does not imply that $x$ is distant from $y$.
Conversely, the \emph{stance} indicator had a higher negative pair score (73.3\%), suggesting that this indicator would perform better in finding negative samples.

\spara{SciLander has the potential to be extended to general domains.}
We based SciLander on features that work as indicators of source similarity or dissimilarity, motivated by previous research on language and misinformation \cite{chambers2018handbook, DBLP:conf/icwsm/HorneNA19, DBLP:conf/www/Smeros0A19}.
The \emph{shift} and \emph{copy} indicators are agnostic to the news domain since they only require the presence of text.
However, the \emph{stance} and \emph{jargon} indicators are closely tied to scientific news.

To extend the application of SciLander to other, including non-scientific, domains, a topic of choice must be specified prior to the application of the method.
The chosen topic must include a set of entities referred to by news sources, such as political figures in the political news domain.
In this case, the stance towards scientific references would be replaced by the stance towards such political figures, and the scientific jargon would be replaced by political jargon.
Topics can be manually defined via a set of keywords and entities, or automatically defined, such as by applying topic modeling to extract the relevant keywords from the news documents.

\spara{Limitations.}
Our methodology was only applied and evaluated on an English dataset; extending to other languages would only require translation/adaptation of the domain-specific lexicon used to compute the jargon indicator or simply skipping this indicator and training using the other three.
All the other indicators as well as the introduced embedding model, are based either on language-agnostic or already multilingual models.

Furthermore, our methodology supports only explicit citations, i.e., direct outgoing links to scientific papers, and not implicit mentions of science-related entities (e.g., universities) because the latter design choice introduces ambiguity and noisy source triplets.

Finally, we implicitly filter the scientific references related to COVID-19 as we filter the news corpus citing these references. 
Explicit filtering would require downloading and parsing the references from different formats, e.g., \emph{pdf}, which is a demanding task not in the scope of this work.

\section{Conclusions}\label{sec:conclusions}

We have introduced SciLander, a method for learning a representation of news sources reporting science-related content. 
Our method uses a combination of signals to estimate the similarity between news sources. 
We have shown that these signals complement each other, capturing relationships between distinct sets of sources from a dataset of news articles related to COVID-19. 
Furthermore, the features learned by our model demonstrated superior performance to baselines for the task of source credibility detection, both in an offline and an online setting, requiring as little as three months of publication activity to accurately classify news sources.
Lastly, we have shown that the learned source representations encode information of credibility and political leaning, forming clusters of sources that show similar reliability and political bias.
In particular, we discovered two large clusters of unreliable sources to which different types of conspiracy news sources flock. 
One of them concentrates on alternative health misinformation, and the other promotes hyper-partisan political conspiracies.

\spara{Reproducibility.}
All the data, code, and models used for this paper are publicly available for research purposes in the following repository: \textbf{\url{https://github.com/mgruppi/SciLander}}.

\section*{Ethics Statement}

Our work aims at finding representations that capture the similarities
and differences between news sources in their coverage of the
COVID-19 pandemic. 
Our proposed method bases this representation on the language usage, content copy/sharing behavior, and their stance towards scientific references.
We show that the representations learned from these signals are useful for several downstream tasks, including understanding the reliability of a source.
This is accomplished by using proxies to trust scientific references, language, and content. 
One must be careful when applying this method to untested dimensions, such as the presence of language usage by minority groups. 
These groups may be underrepresented in the training data, which may cause the model to make biased predictions about them. We propose that this method aids the decision-making process as a complement to human judgment rather than a replacement.

\section*{Acknowledgements}
This work was supported by the Rensselaer-IBM AI Research Collaboration (\url{http://airc.rpi.edu}), part of the IBM AI Horizons Network (\url{http://ibm.biz/AIHorizons}).

\clearpage
\balance
\bibliography{references-sm}

\end{document}